\documentclass[prb,twocolumn,superscriptaddress,showpacs,amsmath,amssymb]{revtex4-2}
\usepackage{natbib}
\usepackage{float}
\usepackage{graphicx}%

\usepackage{dcolumn}
\usepackage{bm}
\usepackage{latexsym,epsfig}
\usepackage{graphicx}
\usepackage{verbatim}
\usepackage{comment}
\usepackage{amsmath}
\usepackage[utf8]{inputenc}
\usepackage{amssymb}
\usepackage{physics}
\usepackage{stmaryrd}
\usepackage{color}
\usepackage{epstopdf}
\usepackage{grffile}
\usepackage{lipsum}
\usepackage{enumitem}

\usepackage[dvipsnames]{xcolor}
\DeclareGraphicsExtensions{.eps}

\newcommand{\beq}{\begin{equation}}
\newcommand{\eeq}{\end{equation}}
\newcommand{\bea}{\begin{eqnarray}}
\newcommand{\eea}{\end{eqnarray}}
\newcommand{\ben}{\begin{eqnarray*}}
\newcommand{\een}{\end{eqnarray*}}
\newcommand{\bfig}{\begin{figure}}
\newcommand{\efig}{\end{figure}}
\usepackage{soul}
\usepackage{hyperref}
\usepackage{cancel}
\hypersetup{
    colorlinks=true,      
    urlcolor=blue,
    citecolor=blue,
    linkcolor=blue
}

\usepackage{amsmath,amssymb}
\usepackage{bm}
\usepackage{graphicx}
\usepackage{xcolor}
\usepackage{hyperref}
\usepackage{physics}
\usepackage{booktabs}
\usepackage{dcolumn}

\hypersetup{colorlinks=true,linkcolor=blue!60!black,
            citecolor=blue!60!black,urlcolor=blue!60!black}

\begin{document}

\title{Many-Body Mobility Edge and Non-Hermitian Skin Effect in an Interacting Quasi-Periodic Spin Chain}

\author{Lavoisier Wah}
\email[]{lavoisier.wahkenounouh@mpl.mpg.de}
\affiliation{Max Planck Institute for the Science of Light, 91058 Erlangen, Germany}
\affiliation{Department of Physics, Friedrich-Alexander-Universit\"at Erlangen-N\"urnberg, 91058 Erlangen, Germany}

\author{Ayan Banerjee}
\email[]{ayan.banerjee@mpl.mpg.de}
\affiliation{Max Planck Institute for the Science of Light, 91058 Erlangen, Germany}

\author{Flore K. Kunst}
\email[]{flore.kunst@mpl.mpg.de}
\affiliation{Max Planck Institute for the Science of Light, 91058 Erlangen, Germany}
\affiliation{Department of Physics, Friedrich-Alexander-Universit\"at Erlangen-N\"urnberg, 91058 Erlangen, Germany}

\date{\today}

\begin{abstract}

Non-Hermitian many-body physics reveals a rich interplay between topology, localization, and boundary effects, yet their collective behavior in interacting disordered systems remains largely unexplored. In this work, we study an interacting non-Hermitian spin chain subject to a quasi-periodic longitudinal field, providing a unified and controlled setting, where non-Hermitian dynamics, interactions, and localization mechanisms intertwine. Remarkably, we discover a \textit{``D‑shaped''} many-body mobility edge that separates extended and localized eigenstates, while simultaneously delineating regimes of many-body localization and the many-body skin effect—where many-body eigenstates acquire an anomalous drift towards a boundary under open boundaries—emerging from the combined action of interactions, non-Hermiticity, and driving amplitude. We demonstrate that the skin effect induces multifractal scaling in the non-Hermitian eigenstates, providing a clear signature of the many-body skin effect. Employing diagnostics such as the fractal dimension, complex eigenvalue fractions, and many‑body inverse participation ratios, we map out a unified phase diagram in which all measures consistently identify the \textit{``D‑shaped''} mobility edge. Finally, we probe this interplay using both complex level-spacing statistics and dynamical observablessuch as density imbalance, entanglement growth, and wave-packet evolution, culminating in a rich many-body mobility phase diagram that captures both the many-body skin effect and localization transitions. Our results identify a clear, defining signature of the \textit{``D-shaped''} many-body mobility edge, and underscore its pivotal role in shaping the physics of open quantum many-body systems.
\end{abstract}

\maketitle

\section{Introduction}\label{S0}

Quantum many-body systems with disorder and interactions can either thermalize or remain localized, where the latter phenomenon is known as many-body localization (MBL)~\cite{PhysRevB.21.2366,altman2018many,PhysRevLett.113.107204,bairey2017driving, RevModPhys.91.021001}. A crucial aspect of MBL and related phenomena is the concept of a mobility edge, which separates localized from extended states in energy space and governs the transition between insulating and conducting behavior in quantum systems~\cite{PhysRevB.91.081103,PhysRevLett.113.200405,PhysRevB.107.045108,PhysRevB.102.014310,10.21468/SciPostPhys.1.1.010}. Its presence in interacting or driven systems challenges conventional localization theory and signals the emergence of rich non-extended phases in quantum condensed matter~\cite{PhysRevB.91.081103,PhysRevB.92.064203,PhysRevResearch.2.032045}. In single-particle physics, mobility edges emerge in three dimensions, two-dimensional systems with broken time-reversal symmetry (such as the integer quantum Hall effect), and quasi-periodic potentials, but their existence in interacting systems remains debated~\cite{PhysRev.109.1492,PhysRevB.25.2185,RevModPhys.67.357,PhysRevLett.104.070601,PhysRevB.93.014203,PhysRevB.107.035129,PhysRevB.97.104204,PhysRevLett.114.146601,PhysRevB.95.155129}.

Remarkably, recent numerical work revealed that a single-particle mobility edge in an incommensurate potential can survive in the many-body spectrum, giving rise to a tunable many-body localization edge or many-body mobility edge (MBME)~\cite{PhysRevLett.115.230401,PhysRevLett.115.186601,PhysRevLett.122.170403,PhysRevB.93.184204}. This opens a pathway to explore energy-resolved thermalization and its breakdown via diagnostics like level statistics and entanglement entropy~\cite{PhysRevLett.113.107204,PhysRevA.103.023323,deng2017many,PhysRevB.92.195153}. Furthermore, multifractality has emerged as a unifying concept across diverse areas of condensed matter physics, capturing the intricate spatial and spectral fluctuations of quantum states in complex systems~\cite{RevModPhys.80.1355}. In many-body settings, multifractal structures arise from the interplay between disorder, interactions, and quantum interference, leading to highly nontrivial scaling of wave function amplitudes in Hilbert space~\cite{PhysRevB.91.081103,PhysRevB.96.104201}. Near localization transitions, such as those between ergodic and many-body localized phases, multifractality provides a quantitative framework for characterizing criticality beyond conventional universality classes~\cite{PhysRev.109.1492,PhysRevLett.42.673,PhysRevLett.123.180601}. Moreover, recent developments highlight its relevance in driven, open, and monitored quantum systems, where the competition between coherence, dissipation, and measurement reshapes the multifractal landscape~\cite{PhysRevLett.128.050602,PhysRevB.104.214307,PhysRevLett.128.130605}.

At the same time, the search for topological properties in non‑Hermitian (NH) systems has accelerated rapidly over the past few years, emerging as a vibrant research front at the intersection of condensed matter physics and optics~\cite{RevModPhys.93.015005,okuma2023non,banerjee2023non}. Recent developments in NH systems have revealed a rich tapestry of phenomena. Among the most striking, is the non-Hermitian skin effect~\cite{zhang2022review,gohsrich2025non,okuma2020topological,PhysRevLett.121.086803}, where an extensive number of bulk eigenstates become exponentially localized at the boundaries, necessitating a refined bulk–boundary correspondence \cite{PhysRevLett.121.026808, gohsrich2025non,PhysRevLett.120.146402}. When interactions and disorder (correlated or random) are introduced into NH systems, specific models have been shown to exhibit MBL transitions accompanied by real-to-complex spectral transitions and, under open boundaries, the emergence of the many-body skin effect~(MBSE), both in static~\cite{PhysRevLett.123.090603,PhysRevB.108.184205,roccati2024diagnosing} and driven settings~\cite{banerjee2025multiple}.

Furthermore, the interplay between multifractality and the skin effect has emerged as an active frontier of research: The suppression of the multifractal dimension under open boundary conditions~(OBCs) relative to the periodic case reveals that the MBSE has a direct manifestation in Fock space, called the Fock-space skin effect~\cite{Shimomura2024}, whereby right eigenstates accumulate on a sub-extensive corner of the many-body Hilbert space, generally distinct from both conventional real-space localization and ergodic delocalization~\cite{Hamanaka2025}.
The study of the interplay between single‑particle mobility edges and NH spectral topology has surged in recent years, driven by theoretical and experimental advances in quasi-periodic systems~\cite{li2024ring,xia2022exact,liu2020generalized,liu2021exact}. In sharp contrast, the elusive nature of MBMEs in many-body systems poses a significant challenge particularly in NH setups~\cite{PhysRevB.110.165101}, where the MBME's behavior remains largely unexplored. The investigation of the MBME thus forms the very raison d'\^etre of the present study. In addition, we shall discuss the emergence of the MBSE and MBL.

To achieve these goals, we study a one-dimensional interacting NH spin chain subjected to a quasiperiodic longitudinal field. The model is introduced and its many-body energy spectra are
characterized in Sec.~\ref{S1}. In Sec.~\ref{S2}, we demonstrate that the system hosts a D-shaped MBME and exhibits multifractal eigenstates, diagnosed through $\langle D_2\rangle$ and the complex spacing ratio $\langle r\rangle$. In Sec.~\ref{S4}, we construct the unified phase diagram in the $(J,h)$ plane, explore representative cuts along the $J$, $h$, and
$\gamma$ axes to isolate the role of each parameter, and characterize the real-to-complex spectral phase transition. In Sec.~\ref{S5}, we investigate dynamical observables, establishing their connection to the phase diagram and the MBME.
Finally, Sec.~\ref{S6} summarizes our findings and presents an outlook.

\begin{figure}
    \centering
  \includegraphics[width=0.475\textwidth]{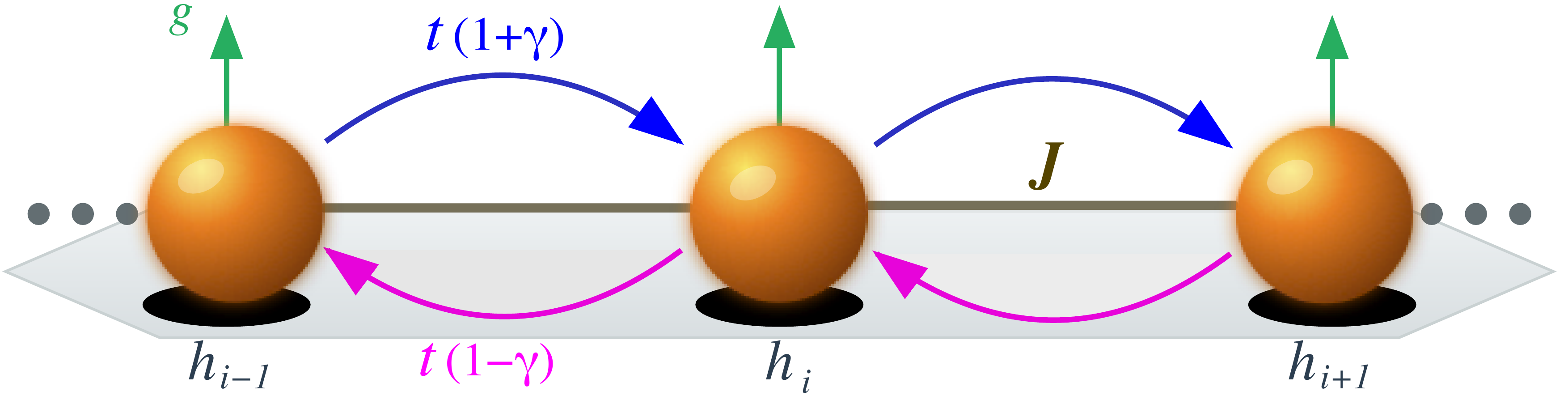}
    \caption{\textbf{Interacting NH spin chain}. Schematic representation of the NH spin chain with a longitudinal quasi-periodic field in Eqs.~\eqref{Eq1} and \eqref{Eq2}. The system includes asymmetric hoppings captured by $t$ and $\gamma$, nearest-neighbor $zz$-interactions ($J$), a transverse field ($g$), and an on-site longitudinal field ($h_i$) given Eq.~\eqref{Eq:QP}, which plays the role of disorder. }
    \label{fig1}
\end{figure}

\section{Non-Hermitian spin chain with quasi-periodic longitudinal field}\label{S1}

\subsection{Model Hamiltonian }

We consider an interacting non-Hermitian spin chain of length $L$ subject 
to a quasi-periodic longitudinal potential (see Fig.~\ref{fig1}). The Hamiltonian 
is given by
\begin{equation}\label{Eq1}
    H= \sum_{i=1}^L \left( H_t +J\sigma_i^z\sigma_{i+1}^z + g\sigma_i^x +h_i\sigma_i^z \right),
\end{equation}
where the hopping Hamiltonian $H_t$ is
\begin{align}\label{Eq2}
   H_t = t\left[(1+\gamma)\sigma_i^-\sigma_{i+1}^+ + (1-\gamma)\sigma_i^+\sigma_{i+1}^-\right]
\end{align}
with $t \in \mathbb{R}$ the hopping amplitude, $J \in \mathbb{R}$ the 
nearest-neighbor Ising ($zz$) coupling, $g \in \mathbb{R}$ the transverse 
field, and $\gamma \in \mathbb{R}$ the non-Hermitian asymmetric hopping 
parameter. The operators $\sigma_i^{x,y,z}$ denote the Pauli matrices at 
site $i$, with $\sigma_i^{\pm} := \sigma_i^x \pm i\sigma_i^y$ the spin 
raising and lowering operators, respectively. 

Unlike Ref.~\cite{Hamanaka2025}, which studies the many-body skin effect and its multifractal structure in a clean non-Hermitian interacting system with $h_i = h$, the central ingredient of our model is the site-dependent longitudinal field $h_i$. This field encodes a quasi-periodic (Aubry-Andr\'{e}-type) potential
\begin{align}\label{Eq:QP}
    h_i = h\cos\!\left(2\pi\alpha\, i + \phi\right),
\end{align}
where $h \in \mathbb{R}$ controls the disorder strength, $\alpha$ is an 
irrational number (typically chosen as the inverse golden ratio 
$\alpha = (\sqrt{5}-1)/2$ to ensure incommensurability with the lattice), 
and $\phi \in [0, 2\pi)$ is an arbitrary phase offset.
The quasi-periodic modulation of $h_i$ introduces a competition between 
non-Hermitian effects, many-body interactions, and disorder-driven 
localization, which is the central focus of the present work.

\begin{figure*}[ht!]
    \includegraphics[width=0.854\textwidth]{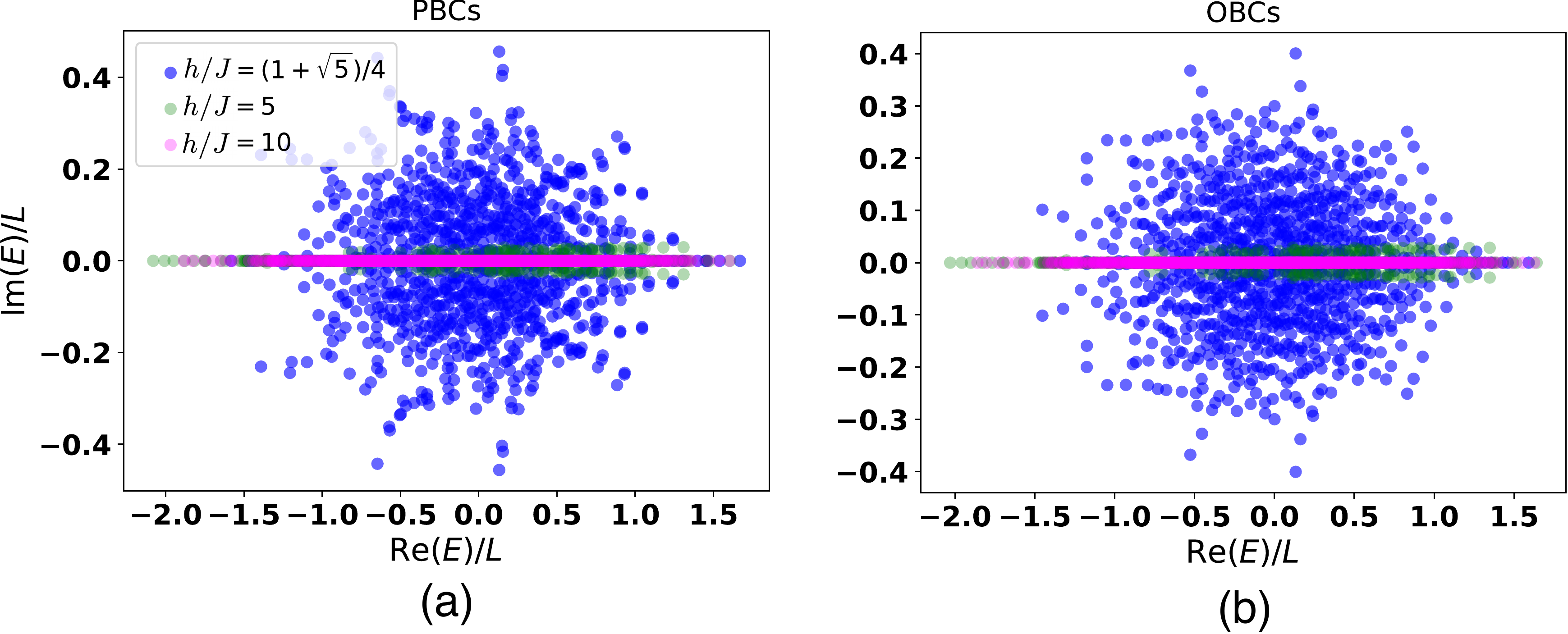}
    \caption{\textbf{Energy spectrum}. We plot the spectrum per spin of the model in Eqs.~\eqref{Eq1}-\eqref{Eq:QP} under \textbf{(a)} PBCs and \textbf{(b)} OBCs on the complex plane for $h/J = (1+\sqrt{5})/4$ (blue), $h/J = 5$ (green), and $h/J=10$ (magenta). We consider $L=10$, $J=1$, $t/J=1/\sqrt{2}$, $\gamma/J=0.8$, and $g/J=(5+\sqrt{5})/8$. We keep these values fixed unless otherwise specified.}
    \label{fig2}%
\end{figure*}

Let us discuss the role of the different parameters in our Hamiltonian. For $\gamma = 0$ the system reduces to the standard Hermitian XXZ spin chain with both transverse and longitudinal fields~\cite{PhysRevB.77.064426, PhysRevB.82.174411}. This paradigmatic interacting system can thermalize or exhibit MBL depending on the quasi-periodic potential $h_i$~\cite{PhysRevB.87.134202}. Turning the NH asymmetry parameter $\gamma$ on, we see from Eq.~\eqref{Eq2} that this term controls the imbalance in the spin-flip hopping, giving rise to asymmetric spin transport. Physically, such a non-reciprocal hopping can be realized via continuous monitoring with post-selection on null measurement outcomes~\cite{RevModPhys.70.101, Daley2014QuantumTA}, and bears a direct connection to the asymmetric simple exclusion process~\cite{DERRIDA199865,PhysRevLett.77.570}. In the non-interacting limit, $J = g = h_i = 0$, the Hamiltonian reduces to the Hatano-Nelson model \cite{PhysRevLett.77.570}. This limit exhibits the celebrated non-Hermitian skin effect. Under periodic boundary conditions (PBCs), all single-particle states are extended Bloch waves, while under OBCs, they become exponentially localized at one edge of the chain. Finally, the model in Eq.~\eqref{Eq1} interpolates between these rich limiting cases: it combines non-Hermitian asymmetric hopping, quantum interactions, and quasi-periodic inhomogeneity in a minimal setting, enabling the exploration of the interplay between the MBSE, MBL, and delocalization along with the energy-resolved phase map encoded in the MBME.
As we are interested in studying the MBME, the MBSE and MBL in this work, we set $J = 1$, $t/J=1/\sqrt{2}$, $\gamma/J = 0.8$, and $g/J= (5+\sqrt{5})/8$~\footnote{The transverse field $g/J = (5+\sqrt{5})/8$ is chosen to be algebraically 
related to the golden ratio $(1+\sqrt{5})/2$, ensuring that it shares the same irrationality as the quasiperiodic modulation parameter 
$\alpha = (\sqrt{5}-1)/2$, thereby avoiding accidental commensurability with the lattice and preventing fine-tuned resonances in the spectrum.}, unless otherwise specified, and study the phenomenology of our model when varying $h/J$.

\subsection{Energy spectrum}

We begin by characterizing the energy spectra across distinct regimes of the quasi-periodic amplitude $h$. For small disorder [$h=(1+\sqrt{5})/4 << 10 $], the NH hoping terms dominate the system's dynamics over the longitudinal field, and give rise to an asymmetric transport of spins in the system. As a consequence, the energy spectrum is complex under both PBCs and OBCs, cf. the blue dots in Fig.~\ref{fig2}. Note that our model respects time-reversal
symmetry (that is, it is invariant to complex conjugation), which also requires the complex spectrum to be symmetric about the real axis both for PBCs and OBCs. Indeed, $\gamma$ introduces spatial directionality into the hopping, driving asymmetric spin transport and giving rise to the skin effect. Furthermore, for strong disorder ($h = 10$, magenta in Fig.~\ref{fig2}), the longitudinal field term dominates over both the hopping and interaction terms, and the many-body eigenstates become approximate product states in the $\sigma^z$ basis with each spin aligning along $\pm\hat{z}$ according to the local sign of $h_i = h\cos(2\pi\alpha i + \phi)$. In this regime, the Hamiltonian is approximately diagonal in the $\sigma^z$ basis, with the asymmetric hopping (controlled by $t$ and $\gamma$), the Ising interaction $J$, and the transverse field $g$ all acting as small off-diagonal perturbations. Consequently, the imaginary parts of the eigenvalues are strongly suppressed and the spectrum becomes approximately real, as shown in magenta in Fig.~\ref{fig2}. We see that for the intermediate regime, $h = 5$, (green in Fig.~\ref{fig2}), the spectrum tends towards the real axis but is still partially imaginary.

In the context of OBCs, we anticipate the emergence of the MBSE in weak fields $h$, i.e., in the regime where $\gamma$ dominates over $h$. Specifically, the unidirectional transport of excitations leads to the accumulation of many-body wavefunction amplitudes near a single boundary. This results in the formation of boundary-localized eigenstates, a defining feature of the MBSE, as explored in recent studies \cite{Wang2023, Hamanaka2025, qin2025dynamical}, which we will delve in the forthcoming sections. 

This pronounced boundary sensitivity of the many-body eigenstates (in the presence of weak field) and the nontrivial interplay between non-Hermiticity, interaction and (quasi-periodic) disorder naturally raises the question of how the internal structure of these states behaves across the spectrum. In particular, we ask whether the eigenstates exhibit signatures of criticality such as multifractality (reflecting a phase that is neither fully localized nor fully extended) or whether the system hosts a MBME, which is an energy-dependent transition between localized and extended eigenstates. In what follows, we elucidate these questions by investigating the spatial statistics and scaling properties of the many-body wavefunctions aiming to uncover signatures of multifractality, the MBME, and the MBSE.

\section{Multifractality and Many-body Mobility edge}\label{S2}

\subsection{Multifractal dimension}
In interacting many-body systems, many-body effects modify the nature of Anderson localization, leading to the emergence of MBL \cite{Abanin2019many, Gornyi2005interacting}. During MBL transitions, the intricate structure of the many-body Hilbert space may give rise to a scenario, where the states are neither fully localized nor delocalized, this is known as multifractality \cite{Hamanaka2025}. Indeed, in this context the wavefunctions of the system exhibit a non-trivial (hierarchical, self-similar structure) distribution within the Hilbert space, characterized by a continuous set of exponents and fractal dimensions, rather than a single, constant dimension \cite{Hamanaka2025}. This contrasts with the MBSE, where all wavefunctions are localized at the boundary, and with MBL, which occurs in the absence of ergodicity \cite{PhysRevB.106.064208}. In this section, we propose to investigate the multifractality of the system in Eqs.~\eqref{Eq1}-\eqref{Eq:QP} by studying the role played by the quasi-periodic field.

To quantify the multifractality, we study the multifractal dimension $D_q$, characterizing the scaling behavior of wavefunction amplitudes \cite{Nicolas2019multi}. 
In general, the multifractal dimension $D_q$ can be expressed as
\begin{align} \label{Eq3}
    D_q =\frac{S_q}{\ln \mathcal{N}}, 
\end{align}
with $S_q$ the $q^{th}$ participation entropy, where $q$ are moments also known as the order of the participation entropy, defined as
\begin{align}\label{Eq4}
    S_q = \frac{1}{1-q} \ln \left( \sum_{n=1}^{\mathcal{N}} |\psi_n|^{2q}\right).
\end{align}
Here $\mathcal{N} = 2^L$ is the dimension of the Hilbert space, and $\psi_n$ are the normalized right eigenstates of our system in a given computational basis $|n\rangle$ (expanded in the tensor-product $\sigma^z$ basis) obtained via exact diagonalization that we implemented with Jax \cite{jax2018github}. In this work, we focus on multifractality defined by the right eigenstates~\cite{PhysRevLett.134.180405},
as the MBSE manifests specifically through their spatial accumulation at one boundary under OBCs---a signal that is washed out in a biorthogonal (right--left) formulation.
We note, however, that a biorthogonal description is the natural framework for relaxation dynamics and eigenstate thermalization in NH systems, where the time evolution of observables involves both left and right eigenstates through the biorthogonal decomposition of the identity~\cite{PhysRevLett.127.070402,PhysRevLett.125.230604,PhysRevLett.134.180405}.

The system is said to be in the multifractal regime when $0<D_q<1$, while it is localized (extended) in Hilbert space when $D_q=0$ ($D_q=1$). We stress that $D_q$ measures localization in the many-body Hilbert space, not in real space: a small $D_q$ means the eigenstate is concentrated on an exponentially small number of $\sigma^z$ configurations. For MBL eigenstates, which are close to product states in the $\sigma^z$ basis (see Sec.~\ref{S1}), real-space localization implies Hilbert-space localization, making $D_q$ a reliable diagnostic for
the MBL phase~\cite{yousef2023mobility,pawlik2024mobil}. For the MBSE, the connection is different: under OBCs, the non-Hermitian skin effect spatially confines the right eigenstates to one boundary, reducing the number of $\sigma^z$ configurations that contribute significantly to the wavefunction and thereby suppressing $\langle D_q\rangle$ relative to PBCs. The OBCs--PBCs gap in $\langle D_q\rangle$ thus serves as a Hilbert-space fingerprint of the real-space boundary accumulation characteristic of the MBSE~\cite{Hamanaka2025}, providing a quantitative diagnostic that complements direct real-space observables. Here $\langle D_q\rangle$ denotes the mean fractal dimension of all right eigenstates within each window.

\begin{figure}
    \centering
    \includegraphics[width=0.4685\textwidth]{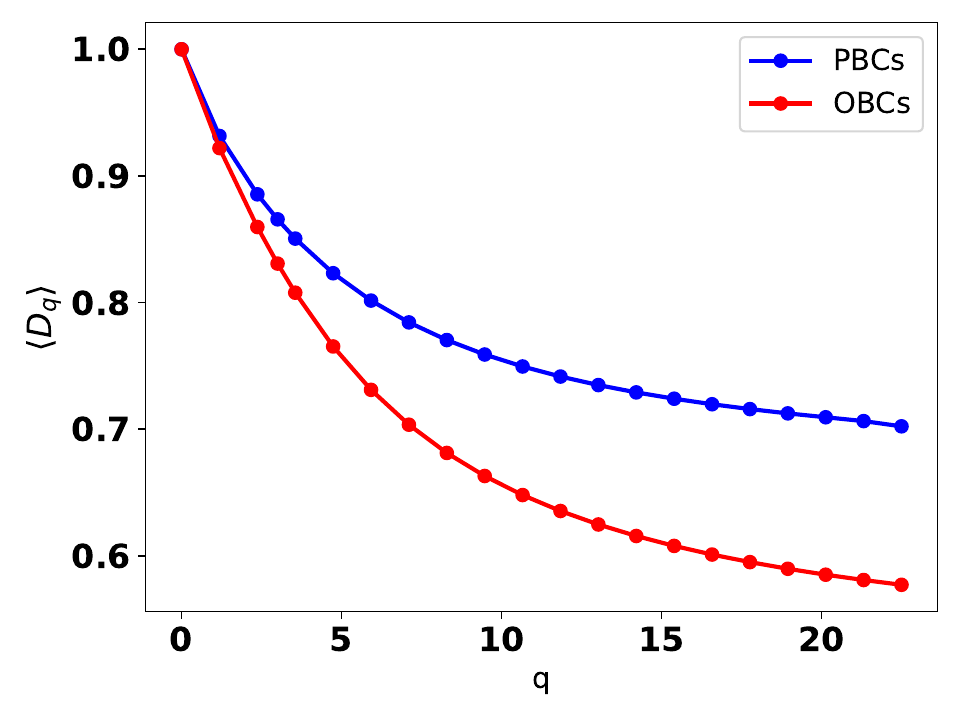}
    \caption{\textbf{Scaling of the mean multifractal dimension $\langle D_q\rangle$}. We plot the average multifractal dimension $\langle D_q\rangle$ in PBCs (blue) and OBCs (red) as a function of $q$ for $L=10$, $J=1$, $t/J=1/\sqrt{2}$, $\gamma/J=0.8$, $h/J = (1+\sqrt{5})/4$, and $g/J=(5+\sqrt{5})/8$. We observe strong multifractality in OBCs as an indicator for the MBSE under OBCs.}
    \label{figA1}
\end{figure}

Computing $D_q$ for our the system, we see that it exhibits strong multifractal behavior under both PBCs (blue in Fig.~\ref{figA1}) and OBCs (red) as $D_q$ varies strongly with varying $q$. The change in $\langle D_q\rangle$ is a signature of multifractality arising from long-normal distributions of wavefunction amplitude \cite{Hamanaka2025,Arnd2019multi}. While both start at $\langle D_q\rangle =1 $ (extended-like at low moment), they diverge at large $q$, where PBCs asymptotes to a greater value than OBCs. This asymmetry reflects the enhanced localization under OBCs, a manifestation of the MBSE. Since we are not interested in any rare events \cite{PhysRevB.110.214210} (e.g., strong localization spots, interference-induced localization, atypical or large wavefunction amplitudes), for numerical stability and in consistancy with the literature, we restrict ourselves to $q=2$ in the following.

\begin{figure*}
    \centering
    \includegraphics[width=1.024\textwidth]{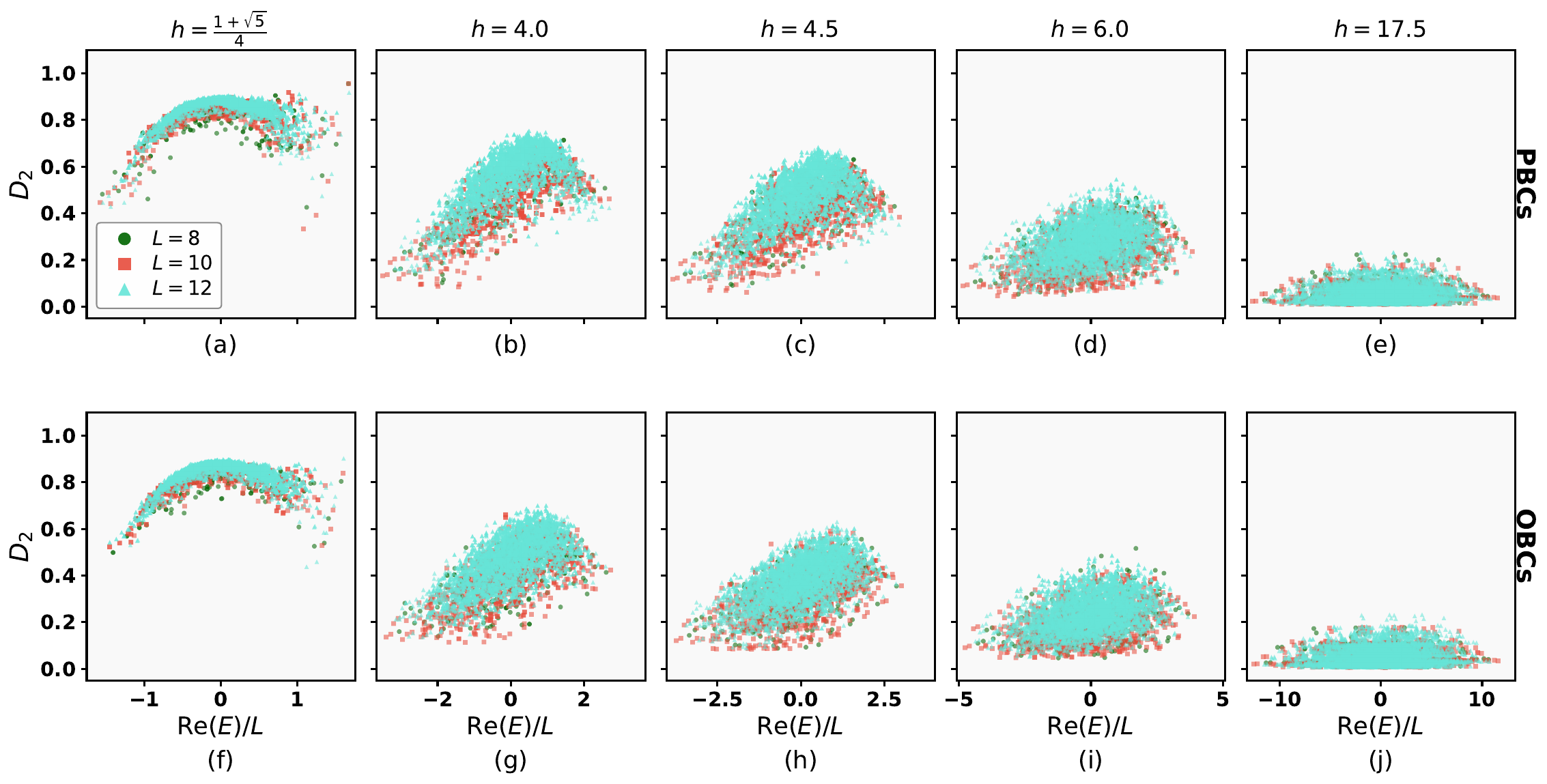}
    \caption{\textbf{Multifractal dimension $\mathbf{D_2}$}. Fractal dimension $D_2$ as a function of the real energy density: $\mathrm{Re}(E)/L$ for system sizes $L=8$ (green), $L=10$ (red), and $L=12$ (cyan), under PBCs~[(a)--(e)] and
OBCs~[(f)--(j)], at five representative values of the quasiperiodic amplitude $h$.
\textbf{(a), (f)}~$h=(1+\sqrt{5})/4$: the extended regime, where $D_2$ is large across the spectrum with a general trend of increasing values for larger system sizes, consistent with flow toward the ergodic fixed point $D_2\to 1$, indicating delocalized eigenstates. \textbf{(b), (g)}~$h=4.0$: proximity to the many-body critical point, consistent with incipient multifractality. \textbf{(c), (h)}~$h=4.5$: the MBME regime, where $\langle D_2\rangle$
increases with $L$ near the spectral center and decreases with $L$ near the spectral edges, revealing the coexistence of extended, critical and MBL eigenstates within the same spectrum, separated by critical energy density. \textbf{(d), (i)}~$h=6.0$: the transition region where the critical window has contracted and $D_2$ is suppressed across the spectrum. \textbf{(e), (j)}~$h=17.5$: the MBL regime, where $D_2\to 0$ uniformly across the spectrum, signaling localized eigenstates concentrated on an exponentially small number of basis states. We set $J=1$, $t/J=1/\sqrt{2}$, $\gamma/J=0.8$, and $g/J=(5+\sqrt{5})/8$.}
    \label{fig3}
\end{figure*}

In Fig.~\ref{fig3}, we plot the spreading of $D_2$ for all $2^L$ eigenstates as a function of the real part of the energy per spin ($\varepsilon=Re(E)/L$) both under PBCs, cf. Fig.~\ref{fig3}(a-e), and OBCs, cf. Fig.~\ref{fig3}(f-j), for different system sizes $L=8,10,12$ (green circles, red squares, and cyan triangles, respectively), and for different values of $h/J$.

For $h= (1+\sqrt{5})/4$ (see Fig.~\ref{fig3}(a, f)), as the system size increases, the $D_2$ values for midspectrum eigenstates ($\varepsilon=0$) logarithmically approach one, which is the expected behavior in the delocalized regime. However, before and after the midspectrum there is no trend towards one, which is a hallmark of multifractality of these eigenstates. This \textit{``crescent-shaped''} spreading of $D_2$ is therefore a signature of multifractality. In addition, under PBCs, $D_2$ is closer to one than under OBCs, which signals the occurrence of the MBSE \cite{Arnd2019multi} as we discuss later. 

The intermediate regime reveals a rich
sequence of spectral phases, shown in Figs.~\ref{fig3}(b)--(d) for PBCs
and Figs.~\ref{fig3}(g)--(i) for OBCs.
At $h=4.0$ [Figs.~\ref{fig3}(b),(g)], $D_2$ takes intermediate values $\in [0.17, 0.65]$
across the entire spectrum for different $L$, and shows no tendency to flow toward either $D_2\to 1$ or $D_2\to 0$ as $L$ increases. This absence of a dominant fixed point suggests proximity to a many-body critical regime, where the competing tendencies of the transverse field $g$, the Ising coupling $J$, and the incommensurate potential are finely balanced.

At $h=4.5$ [Figs.~\ref{fig3}(c),(h)], an energy-resolved structure emerges, providing clear evidence for a MBME accompanied by a significant presence of critical states. Near the spectral centre, larger system sizes sit consistently \textit{above} smaller ones ($D_2\to 0.6$), while near the spectral edges the ordering reverses ($D_2\to 0.2$). This reversal of the finite-size flow direction as a function of $\varepsilon$, as is unambiguously visible in both boundary condition panels, indicates that extended and MBL eigenstates coexist within the same system at the same $h$ separated by a critical energy density, and as such is the defining signature of an MBME~\cite{yousef2023mobility, pawlik2024mobil} (see also Fig.~\ref{fig:D2map}).  

\begin{figure*}[ht!]
    \centering
\includegraphics[width=1.01\textwidth]{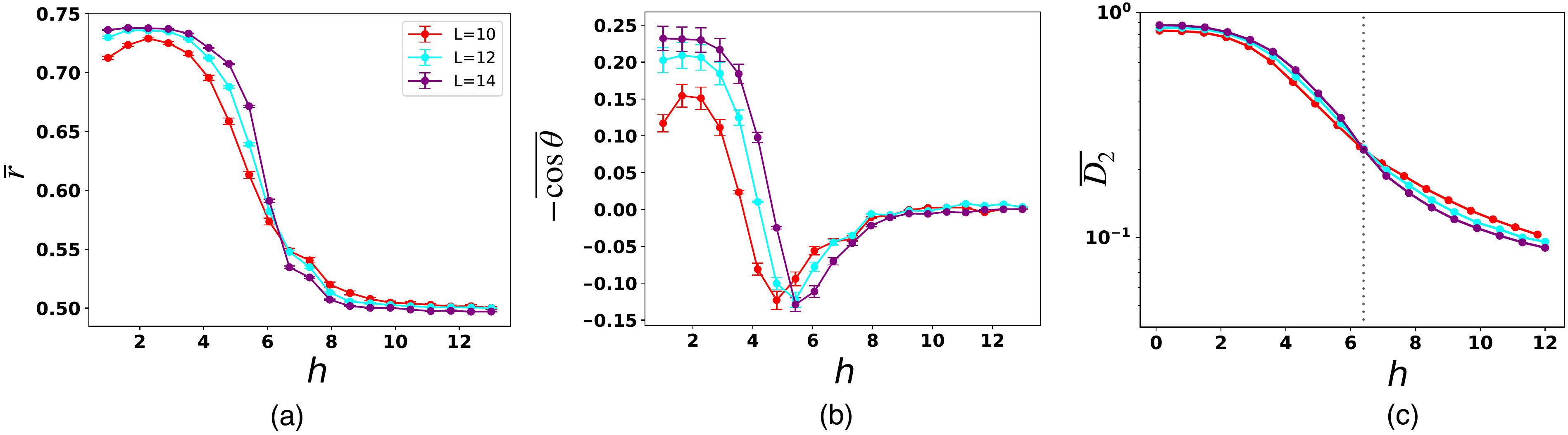}
    \caption{\textbf{Delocalized-to-MBL phase transition}. We plot \textbf{(a)} the mean gap ratio $\bar{r}$,  \textbf{(b)} the angular statistics $-\overline{\cos\theta}$, and \textbf{(c)} the average multifractal dimension $\overline{D_2}$ as a function of the disorder $h$ for different system sizes. One observes that the crossing point of the curves (see \textbf{(a)} and \textbf{(c)}), which is the transition point, falls within the region of $h$, where the MBME is predicted to occur. The universality class of the model changes from Ginibre ensemble statistics in the delocalized regime to Poisson statistics in the localized regime. This suggest a disorder-induced spectral localization transition. We consider $1000, 600$ and $400$ values of $\phi$ for $L=10$, $12$ and $14$, respectively. The transition happens at $h_c^{D_2}\approx 6.5$ (vertical dotted line in (c)). PBCs are considered. We chose $J=1$, $t/J=1/\sqrt{2}$, $\gamma/J=0.8$, and $g/J=(5+\sqrt{5})/8$.}
    \label{fig4}%
\end{figure*}

\begin{figure*}[ht]
\centering
\includegraphics[width=0.93\textwidth]{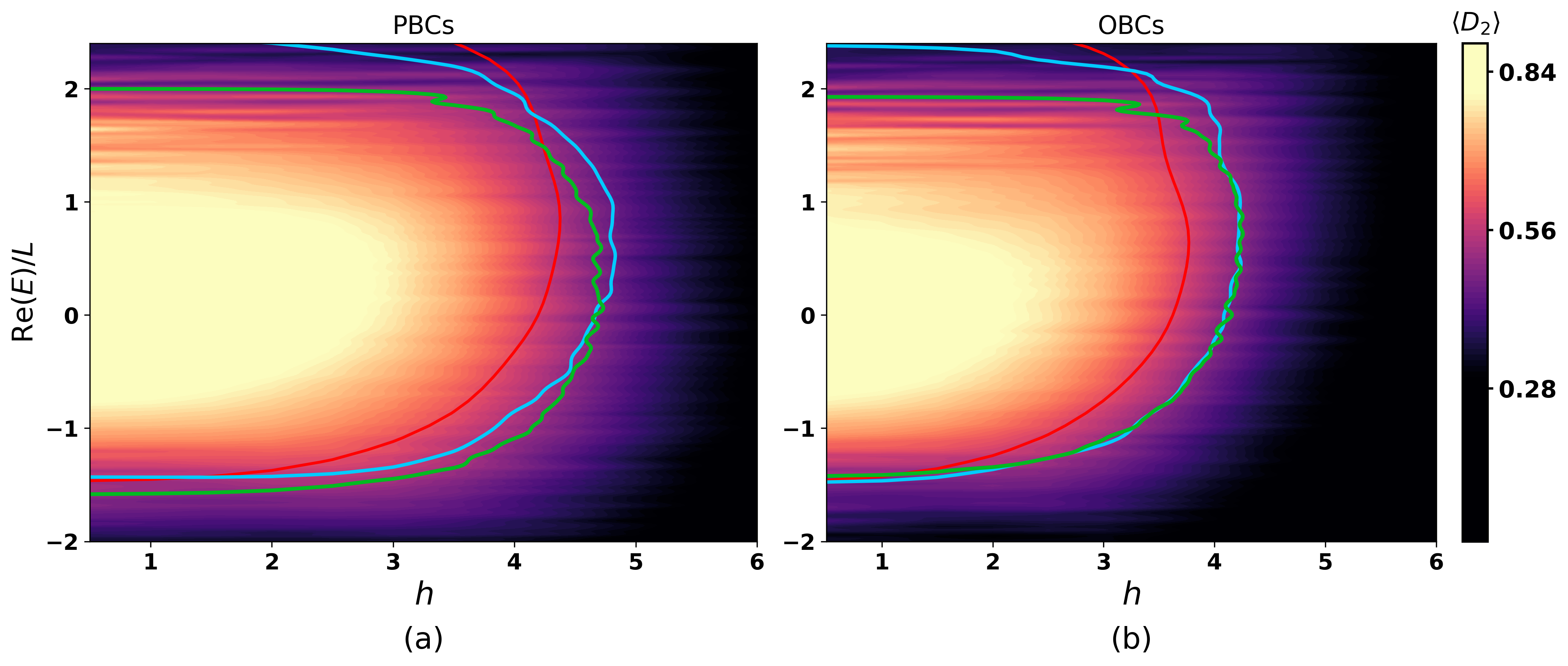}
  \caption{
    \textbf{Energy-resolved phase diagram: The MBME.} We present the fractal dimension $\langle D_2\rangle$ averaged over all eigenstates, whose real energy density $\mathrm{Re}(E)/L$ falls within a uniform spectral window, as a function of $\mathrm{Re}(E)/L$ and the quasiperiodic field $h$ for (a) PBCs and (b) OBCs. The background colour map is computed from the largest system size $L=14$. $\langle D_2\rangle \to 1$ (bright) indicates delocalized eigenstates extended across Hilbert space; $\langle D_2\rangle \to 0$ (dark) marks MBL states. The iso-contours (solid lines) of $\langle D_2\rangle = 0.5$ are shown for $L = 10$ (red),
    $L = 12$ (cyan), and $L = 14$ (green), and trace the MBME separating the
    extended and MBL phases. We set $J=1$, $t/J=1/\sqrt{2}$, $\gamma/J=0.8$, and $g/J=(5+\sqrt{5})/8$.}
\label{fig:D2map}
\end{figure*}

At $h=6.0$ [Figs.~\ref{fig3}(d),(i)], the extended window has contracted: the overall $D_2$ is suppressed across the spectrum, the size-separation near the spectral center is visibly weaker than at $h=4.5$, and the spectral support in $\varepsilon$ has broadened as the quasiperiodic field increasingly controls the many-body bandwidth. These features place $h=6.0$ in the transition region between the MBME phase and a fully MBL spectrum, where the mobility edge has moved inward toward the spectral centre, consistent with the phase boundary extracted from the two-dimensional $\langle D_2\rangle$ map in Fig.~\ref{fig4}(c) and 
Fig.~\ref{fig:D2map}.

Finally, in the strong field regime with $h=17.5$, cf. Fig.~\ref{fig3}(e,j), the lower \textit{``crescent''} flattens to the line $D_2 =0$, while the upper one persists but with less spreading, and with a maximum approaching $D_2 \approx 0.2$ (for PBCs and OBCs). This signals that the majority of the eigenstate, which are at $D_2 =0$, are strongly localized. The upper \textit{``cressent''} are localized states with weak hybridization (where amplitudes decay but still exhibit fine structure) resulting from residual nonreciprocal hopping-induced fluctuations but not true delocalization. Interested readers are referred to Refs.~\cite{Rafi2022crit, yang2025tailoring, Zhang2025obser} for more information about hybridization.

\subsection{Complex level-spacing ratio}

To further demonstrate the presence of the MBME, we plot the complex gap ratio also known as the complex level-spacing ratio \cite{PhysRevX.10.021019}, and we show that there is indeed a phase transition happening at some given values of the (quasi-periodic) disorder. The gap ratio $z$ is a dimensionless complex quantity given by
\begin{align} \label{eq:complex_gap_ratio}
    z_i = \frac{E_i -E_i^{nn}}{E_i -E_i^{nnn}} = r_i e^{i\theta_i},
\end{align}
such that its amplitude is $r_i=|z_i|$, where $i=1,...,\mathcal{N}$, and $E_i^{nn}$ ($E_i^{nnn}$) denotes the nearest neighbor~($nn$) and next-nearest
neighbor~($nnn$), respectively, of $E_i$ in the complex plane, determined by minimizing the Euclidean distance $|E_j - E_i|$ over all $j\neq i$ (and $j\neq nn$ for the $nnn$)~\cite{PhysRevX.10.021019}. Here, $\theta$ represents the phase of the complex gap ratio (angular statistics). For each disorder realization (each value of $\phi$), we compute the complex spacing ratio $z_i$ for every eigenvalue $E_i$ in the bulk of the spectrum and report the mean
\begin{align}
    \bar{r} = \langle r_i \rangle,
\end{align}
averaged over both bulk eigenstates and disorder realizations, using 1000, 600, and 400 values of $\phi$ for $L=10$, $12$, and $14$, respectively, excluding spectral edges (the first and the last site), where finite-size effects are most pronounced. Please note that, the bar denotes averaging over both eigenstates and disorder realizations. The definition in Eq.~\eqref{eq:complex_gap_ratio} generalizes the standard gap ratio of Hermitian 
systems to the complex plane \cite{PhysRevX.10.021019}, and the number of realizations is chosen to ensure statistical convergence across all system sizes.

As shown in Fig.~\ref{fig4}(a), the mean spacing ratio \(\bar{r}\) clearly tracks the (quasi-periodic) disorder-driven spectral transition: at weak disorder it remains fixed at \(\bar{r} \approx 0.74\), matching the Ginibre unitary ensemble~\cite{PhysRevX.10.021019} (which for simplicity we will refer to as Ginibre), then shifts toward \( \bar{r} \approx 0.5\), indicative of real-Poisson statistics in the NH localized phase---distinct from the \(\bar{r} \approx 0.38\) plateau seen in Hermitian MBL. Concurrently, as shown in Fig.~\ref{fig4}(b), the angular statistics -\( \overline{\cos\theta}\), which represents the level repulsion anisotropy in the complex plane, begins at approximately $0.24$ (again consistent with Ginibre predictions), decreases with increasing \(h\), becomes negative, and vanishes (-\(\overline{\cos\theta} = 0\)) at strong disorder, which is consistent with Refs.~\cite{PhysRevX.10.021019,PhysRevB.106.064208}. The joint diagnostics of $\bar r $ and $-\overline{\cos\theta}$, along with the crossing of curves for different system sizes, provide robust spectral signatures of the MBME---marking a transition from a NH delocalized (Ginibre) phase to a localized (real-Poisson) phase. This result is further confirmed by the averaged multifractal dimension $\overline{D_2}$ in Fig.~\ref{fig4}(c). Moreover, we can identify the range of critical values of the field within the MBME consistent with Ref.~\cite{pawlik2024mobil}. The critical points in Fig.~\ref{fig4}(a) and Fig.~\ref{fig4}(c) indeed fall under the MBME region as later discussed in this work. The  crossing of these disorder‑averaged curves is a hallmark of a phase transition, which we interpret as part of the boundary between the extended and localized regimes.

\subsection{D-shaped many-body mobility edge}

The eigenstate spreading shown in Fig.~\ref{fig3}, corroborated by the complex spacing ratio in Fig.~\ref{fig4}, points to a rich energy-resolved phase map; we now make this precise through the two-dimensional phase portrait of $\langle D_2\rangle$. The energy-resolved phase diagram of the model is presented in Fig.~\ref{fig:D2map}, which displays the fractal dimension $\langle D_2\rangle$ as a function $h$ and the real part of the energy density $\varepsilon$, under both PBCs (a) and OBCs (b). The central result of Fig.~\ref{fig:D2map} is a compact, D-shaped island of large $\langle D_2\rangle$, embedded in an otherwise localized background, visible in both boundary-condition panels. The iso-contour $\langle D_2\rangle$, shown for each system size, delineates the MBME in the $(\varepsilon,h)$ plane: at a fixed $h$ within the island, eigenstates near the spectral centre are extended, while those at the spectral edges are many-body localized, with the two phases coexisting within the same Hamiltonian at the same disorder amplitude. This energy-resolved coexistence is the defining hallmark of a spectral MBME~\cite{yousef2023mobility, pawlik2024mobil}. The iso-contour sharpens and its enclosed area grows monotonically from $L=10$ to $L=14$, with a decreasing drift between successive system sizes, consistent with convergence to a sharp thermodynamic MBME rather than its dissolution, in agreement with the behavior found in the quantum sun
model~\cite{pawlik2024mobil}. 

The geometry of the extended island reflects the interplay of two distinct localization tendencies acting along orthogonal axes of the phase diagram.
Along $h$: at small values of the transverse field $g$ and the Ising coupling $J$ sustain extended eigenstates, whereas beyond a critical quasi-priodic field the incommensurate potential localises eigenstates through the many-body Aubry-Andr\'{e} mechanism~\cite{AubryAndre1980}. Along $\varepsilon$: eigenstates near the spectral edges correspond to near-polarized, atypical many-body configurations for which the effective hopping matrix elements are exponentially suppressed, making them more susceptible to localization at any given $h$ compared to mid-spectrum states, where level repulsion and hybridization remain strong. The simultaneous closure of the extended region along both $h$ and $\varepsilon$ produces the observed closed D-shaped boundary. The NH hopping asymmetry induces the skin effect effect, which reshapes spectral weight differently under OBCs and PBCs, the $D$-shaped extended island persists under both (Fig.~\ref{fig:D2map}), confirming the MBME as a robust bulk property. Such a $D$-shaped mobility edge is well established in Hermitian systems~\cite{yousef2023mobility,PhysRevResearch.2.042037,guo2021observation}; its appearance here in a NH interacting chain suggests this feature extends beyond the Hermitian setting.

\section{Interacting phase diagram and Phase transitions}\label{S4}

Having characterized the spectral and multifractal properties of individual eigenstates, we now construct the full phase diagram of the interacting NH chain in the $(J, h)$ plane, employing three complementary diagnostics: the many-body inverse participation ratio (MIPR) $\textit{I}$, cf. Eq.~\eqref{eq:MIPR}, the complex eigenenergies ratio $f$, cf. Eq.~\eqref{eq:complex_energy_ratio}, and the mean fractal dimension $\langle D_2\rangle$. All three observables yield mutually consistent phase boundaries, collectively revealing a D-shaped MBME that separates the extended phase, dominated by the MBSE, from the MBL phase driven by strong quasi-periodic disorder. We further show that the real-to-complex spectral transition, driven by symmetry breaking, coincides with the MBL phase boundary. To interrogate the phase structure, we finally examine several cuts along the $J$, $h$, and $\gamma$ axes, which expose several physically distinct features including the interplay between interactions, disorder, and NH hopping asymmetry. 

\begin{figure*}[ht!]
    \centering
    \includegraphics[width=1.025\textwidth]{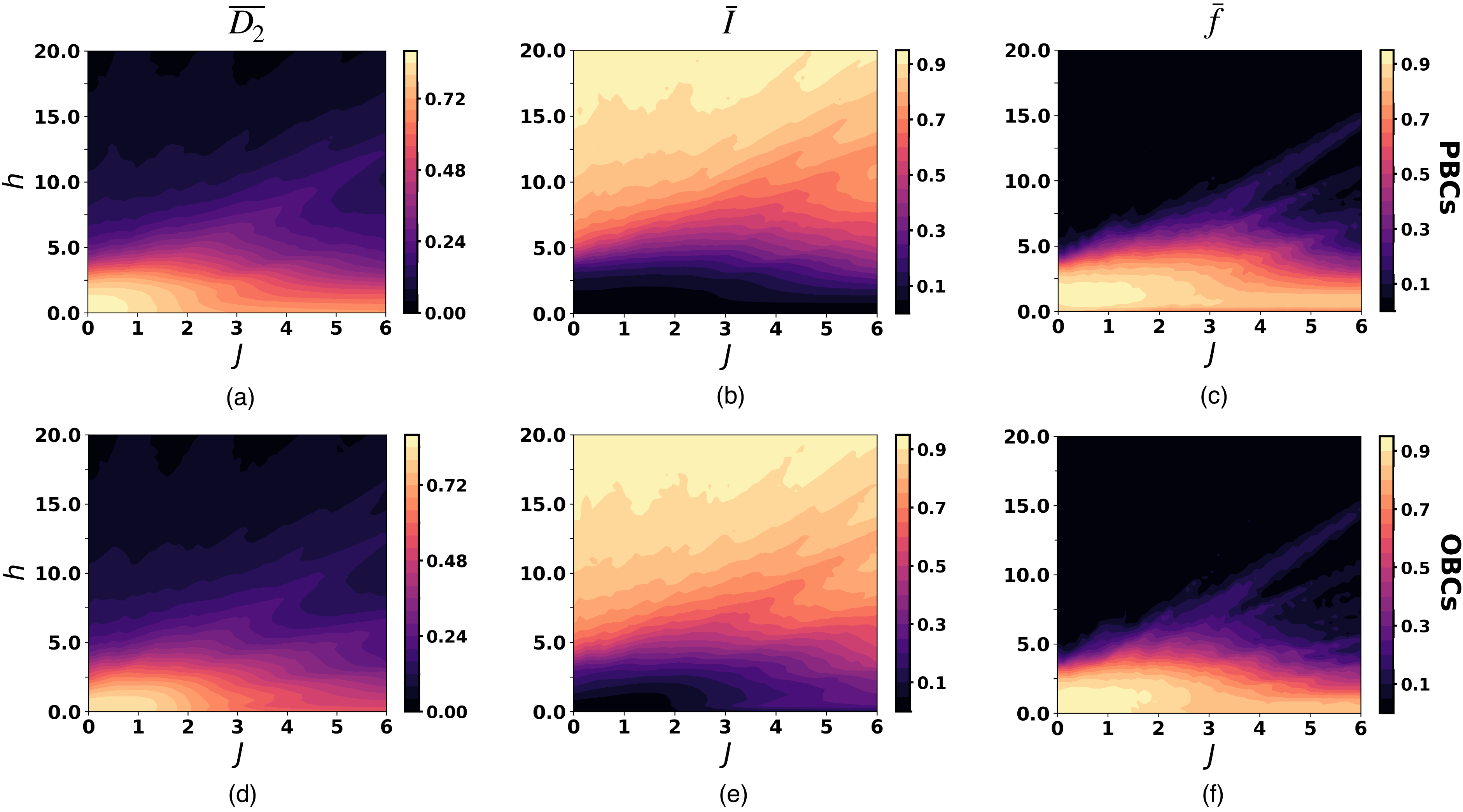}
    \caption{\textbf{Phase diagram}. We plot \textbf{(a,d)} the average multifractal dimension $\overline{D_2}$, \textbf{(b,e)} the many-body inverse participation ratio $\bar{\textit{I}}$, and \textbf{(c, f)} the eigenstate number ration $\bar{f}$ as a function of the nearest-neighbor coupling $J$ and the longitudinal field $h$ under \textbf{(a-c)} PBCs and \textbf{(d-f)} OBCs. We consider $L=10, t=1/\sqrt{2}, \gamma=0.8, g=(5+\sqrt{5})/8 $, and $1000$ values of $\phi$. We observe the extended-to-MBL phase transition with the emergence of the MBME.  }
    \label{fig9}
\end{figure*}

\subsection{Phase diagram}

First, we plot $\overline{D_2}$ as a function of $J$ and $h$ in Fig.~\ref{fig9}(a, d), where we observe that for low interaction and low field amplitide, $\overline{D_2} \approx 0.85$, which suggests that the eigenstates are extended (yellow region). Upon increasing the disorder strength $h$ at fixed interaction $J$, 
the system transitions from the extended phase (yellow) to the MBL phase (dark) through the intermediate \textit{D-shaped} MBME region (purple with $\overline{D_2} \approx 0.5$). Conversely, at fixed $h$, increasing $J$ reveals that stronger interactions promote localization, systematically shifting the MBME toward smaller values of $h$. This interaction-driven 
enhancement of localization is consistent with previous findings in Hermitian settings~\cite{Nandkishore2017many, yousef2023mobility, Deng2020uni}. 

The MIPR $I$ describes how particles are distributed in real space such that when $ I \to 1$, the eigenstates are localized and when $ I \to 0$, the eigenstates are extended~\cite{Ke2023Incomen}. As such, the mean MIPR $\bar{I}$ (averaged over all right eigenstates) characterizes the localization properties of a system. The MIPR is given by
\begin{equation} \label{eq:MIPR}
    \textit{I}= \frac{1}{1-\mu} \left( \frac{1}{\mu L}\sum_{i=1}^L \langle n_i\rangle ^2 -\mu\right),
\end{equation}
where $\mu=0.5$ is the filling factor, and $\langle n_i\rangle$ is the average particle number. We note that restricting the analysis to mid-spectrum eigenstates is not appropriate here, as the phase diagram spans both weak and 
strong interaction regimes~(Fig.~\ref{fig9}(b,e)), where the spectral bulk and edges exhibit qualitatively distinct localization 
properties. As expected, one observe the \textit{``D-shape''} of the MBME (purple region). In contrast to $\overline{D_2}$, the dark region here corresponds to the extended regime, while the yellow region is the MBL regime. Once more, the interaction has the effect of shifting the mobility edge, which is consistent with our previous observations.

Finally, we plot the complex eigenenergies ratio $f$. The complex eigenenergies ratio $f$ measures the variation of the ratio of the complex eigenenergies with nonzero imaginary part \cite{Hamazaki2019NonH, Luitz2015many, zhai2020} and is given by
\begin{equation} \label{eq:complex_energy_ratio}
   f= \left\langle C_i/D \right\rangle, 
\end{equation}
where $C_i$ is the number of eigenenergies with nonzero imaginary part and $D$ the total number of eigenenergies. The cutoff used here is $10^{-13}$ such that for $|\text{Im}(E)|< \text{cutoff}$, the eigenenergy is considered to be real. Again, one observes the same behavior as before. In this case, as for $\overline{D_2}$, $\bar f \to 0$ corresponds to the localized (dark region), $\bar f \to 1$ corresponds to the extended regime (yellow region), and the MBME is within the purple region (see Fig.~\ref{fig9}(c,f)).

\begin{figure*}[ht!]
    \centering
     \includegraphics[width=0.8265\textwidth]{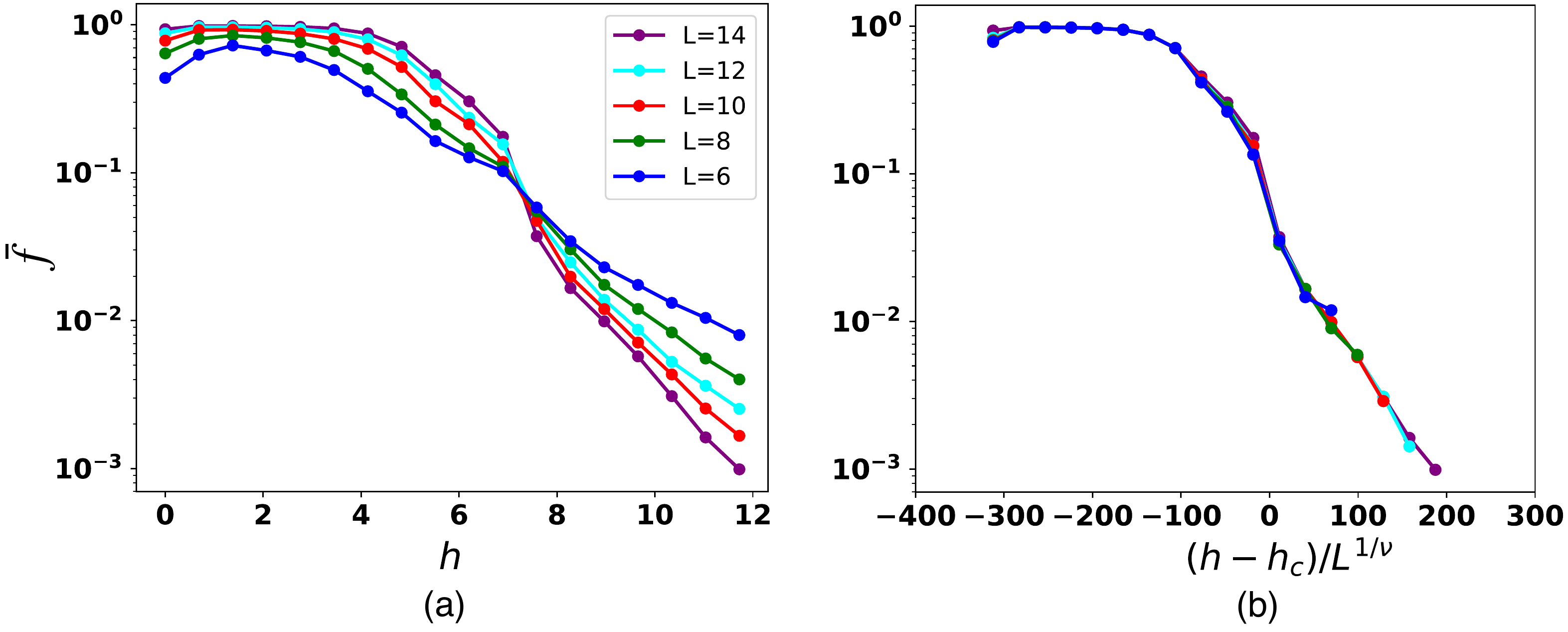}
\caption{\textbf{Phase transition}. We plot \textbf{(a)} the averaged eigenenergies ratio $\bar f$ as a function the longitudinal field strength $h$, and the \textbf{(b)} rescaled curves. We average over $1000$ values of $\phi$ for $L=6$, $8$ and $10$, and 300 values of $\phi$ for $L=12$ and $14$. We consider $t/J=1/\sqrt{2}$, $\gamma/J=0.8$, and $g/J=(5+\sqrt{5})/8 $. One observes the complex-to-real phase transition and the critical scaling collapse of $\bar f$. PBCs are considered.}
    \label{fig10}
\end{figure*}

\subsection{Complex-to-real phase transition}

Beyond the extended-to-MBL transition, the system exhibits a complex-to-real spectral phase transition. The spectrum in our model is symmetric about the real axis [Fig.~\ref{fig2}] due to time-reversal symmetry and an additional spectral symmetry $E(\gamma) = E(-\gamma)$~\footnote{Under time-reversal symmetry, we find $H^*(\gamma) = H(-\gamma)$. Therefore, for the spectrum of $H(\gamma)$ to be symmetric about the real axis, one requires additionally that the spectra of $H(\gamma)$
and $H(-\gamma)$ coincide, which holds here due to the combined symmetry $\mathcal{S}: i \to L+1-i$ (spatial reflection) together with $\gamma\to-\gamma$, as can be verified directly from the Hamiltonian. Together these symmetries guarantee that complex eigenvalues appear
in conjugate pairs $\{E, E^*\}$, rendering the spectrum symmetric about the real axis as observed in Fig.~\ref{fig2}.}. Crucially, strong quasiperiodic disorder suppresses the imaginary parts of the eigenvalues, driving the spectrum from complex to purely real; the complex-to-real transition thereby aligns with the MBL phase transition.

First, we plot the eigenenergies ratio $\bar f$ as a function of the longitudinal field strength for several system sizes $L$ (see Fig.~\ref{fig10}(a)). As the system size increases, $\bar f$ increases for $h \leq h_c \approx 7.1$ and decreases when $h \geq h_c$ indicating a complex-to-real phase transition of the many-body eigenenergies at $h=h_c$ in the thermodynamic limit ($
L\to \infty$). Indeed, when $h < h_c$, the ratio $\bar f \to 1$ meaning that $D_i \approx D$, and as a result almost all the eigenenergies are complex. Likewise, when $h > h_c$, the ratio $\langle f \rangle \to 0$, and $D_i \ll D$ thus almost all the eigenenergies are real. This result then suggests that eigenenergies become almost real with increasing quasi-periodic disorder, which is consistent with the work in Refs.~\cite{Hamazaki2019NonH, zhai2020}.

In addition, the curves of $\bar f$ as a function of $h$ admit a finite-size scaling collapse of the form $\bar f \propto (h-h_c)/L^{1/\nu}$ (see Fig.~\ref{fig10}(b)), yielding a transition point
$h_c\approx 7.1$ and a scaling exponent $\nu=0.7$, consistent with the universality class of quasiperiodic NH systems reported in Ref.~\cite{zhai2020} and distinct from those of random disordered systems~\cite{Hamazaki2019NonH,Luitz2015many}. The transition point $h_c\approx 7.1$ shows a slight upward shift relative to the MBL transition point $h_c^{D_2}\approx 6.5$ extracted from Fig.~\ref{fig4}. We attribute this discrepancy not solely to finite-size effects~\cite{PhysRevLett.123.090603} but also to the fact that the two diagnostics probe genuinely distinct spectral boundaries: $\overline{D_2}$ detects the onset of Hilbert-space localization, while $\bar f$ tracks the fraction of complex eigenvalues, which responds to a different aspect of the complex-to-real crossover~\cite{PhysRevLett.123.090603}. Indeed, Ref.~\cite{PhysRevLett.123.090603} identified two separate boundaries associated with this transition, and the slight mismatch between $h_c$ and $h_c^{D_2}$ is consistent with these two boundaries not coinciding in general.

We note that the fraction $\bar f$ does not reach exactly zero even at large $h$, raising the question of whether true localization, in the sense of a strictly real spectrum, is achieved at the system sizes accessible to exact diagonalization.
This is a genuine open question: a small but nonzero imaginary part of eigenvalues may persist at finite $L$ even deep in the MBL phase, and spectral indicators alone cannot rule out a slowly decaying
$\bar f$ rather than a sharp transition~\cite{crwj-x7j8}. We therefore interpret $h_c$ as a crossover scale at which $\bar f$ begins to decrease with increasing $L$, rather than a sharp thermodynamic transition point, and note that a definitive identification of true MBL in NH systems remains an active area of research.

\subsection{Along different cuts}

\begin{figure*}[t]
    \centering
\includegraphics[width=0.97\textwidth]{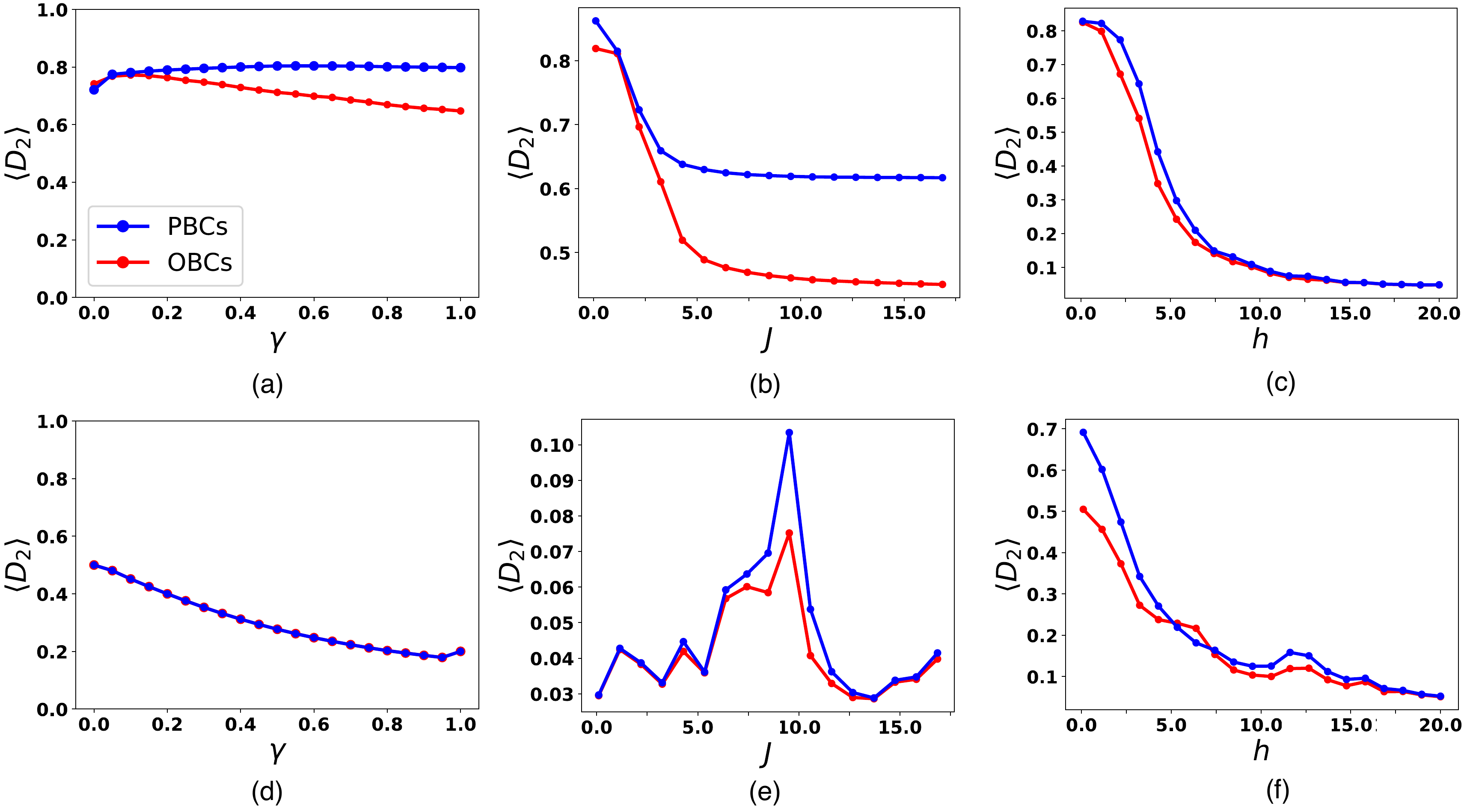}
    \caption{\textbf{Average multifractal dimension}. We plot the average multifractality $\langle D_2 \rangle$ under PBCs (blue) and OBCs (red) as a function of \textbf{(a, d)} the non-Hermitian parameter $\gamma$, \textbf{(b, e)} the nearest-neighbor coupling $J$, and \textbf{(c, f)} the longitudinal field strength $h$. We consider \textbf{(a)} $h=(1+\sqrt{5})/4, J=1.0$,  \textbf{(b)} $h= (1+\sqrt{5})/4, \gamma =0.8$, 
    \textbf{(c)} $J=1.0, \gamma =0.8$, \textbf{(d)} $h=20, J=1.0$  \textbf{(e)} $h=20, \gamma =0.8$, and \textbf{(f)} $J=5.8$, $\gamma =0.8$, for $L=10$. We observe that the presence of the MBSE induces a drop in  $\langle D_2 \rangle$ under OBCs (see \textbf{(a, d)}). Also, for strong disorder, the system becomes boundary insensitive, and both PBCs and OBCs curves superimpose (see \textbf{(c, f)}) suggesting a MBL regime.} 
    \label{fig6}%
\end{figure*}

To further investigate the phase diagram, we examine $\langle D_2\rangle$ averaged over all right eigenstates along representative cuts in the $(J, h, \gamma)$ parameter space, isolating the role of interactions, quasi-periodic disorder, and hopping asymmetry separately.
We note that restricting the average to mid-spectrum eigenstates yields no significant deviation from the full-spectrum result~\cite{Hamanaka2025}, justifying the use of the full eigenstate average throughout. In particular, we highlight the signatures of the MBSE by tracking how $\langle D_2\rangle$ evolves under OBCs and PBCs as each parameter is varied independently.

In Fig.~\ref{fig6}(a) and (d), we plot $\langle D_2\rangle$ as a function of $\gamma$ for weak and strong $h$, respectively, and weak $J$. Particularly, in Fig.~\ref{fig6}(a) one observes that for an increasing NH parameter $\gamma$, $\langle D_2 \rangle$ nearly remains constant under PBCs while decreasing under OBCs suggesting the presence of a strong skin effect in the later. Indeed, under PBCs, the translational symmetry forces any directional bias to cancel out, preventing the accumulation of eigenstates and thus there is no MBSE. However, this translational symmetry is broken under OBCs, exposing the non-reciprocal nature of the system and inducing an accumulation of eigenstates on a privileged boundary (direction). This manifests as a drop in the $\langle D_2 \rangle$ value since the states become less extended and more boundary-localized---this is a distinct signature of the MBSE. Interestingly, for small $h$ under OBCs cf. Fig.~\ref{fig6}(a), $\langle D_2 \rangle$ slowly decreases as $\gamma$ increases suggesting that the the degree of localization grows smoothly with $\gamma$ making them MBSE more pronounced. However, for strong $h$, cf. Fig.~\ref{fig6}(d), $\langle D_2 \rangle$ rapidly decays in both the PBC and OBC case due to the MBL and the system becomes boundary insensitive. In addition, for weak $h$ the PBC and OBC curves have the same starting point cf. Fig.~\ref{fig6}(a), and eventually split alsmost immediately at some particular value of $\gamma$ since the system is initially extended. However, under strong $h$, cf. Fig.~\ref{fig6}(d) the translational symmetry is immediately broken by the quasi-periodic potential. In particular, with increasing $\gamma$, the eigenstates tend to localize, and this localization is exacerbated with the presence of MBL yielding an immediate and fast decrease in $\langle D_2 \rangle$.

\begin{figure*}[ht!]
    \centering
    \includegraphics[width=1.024\textwidth]{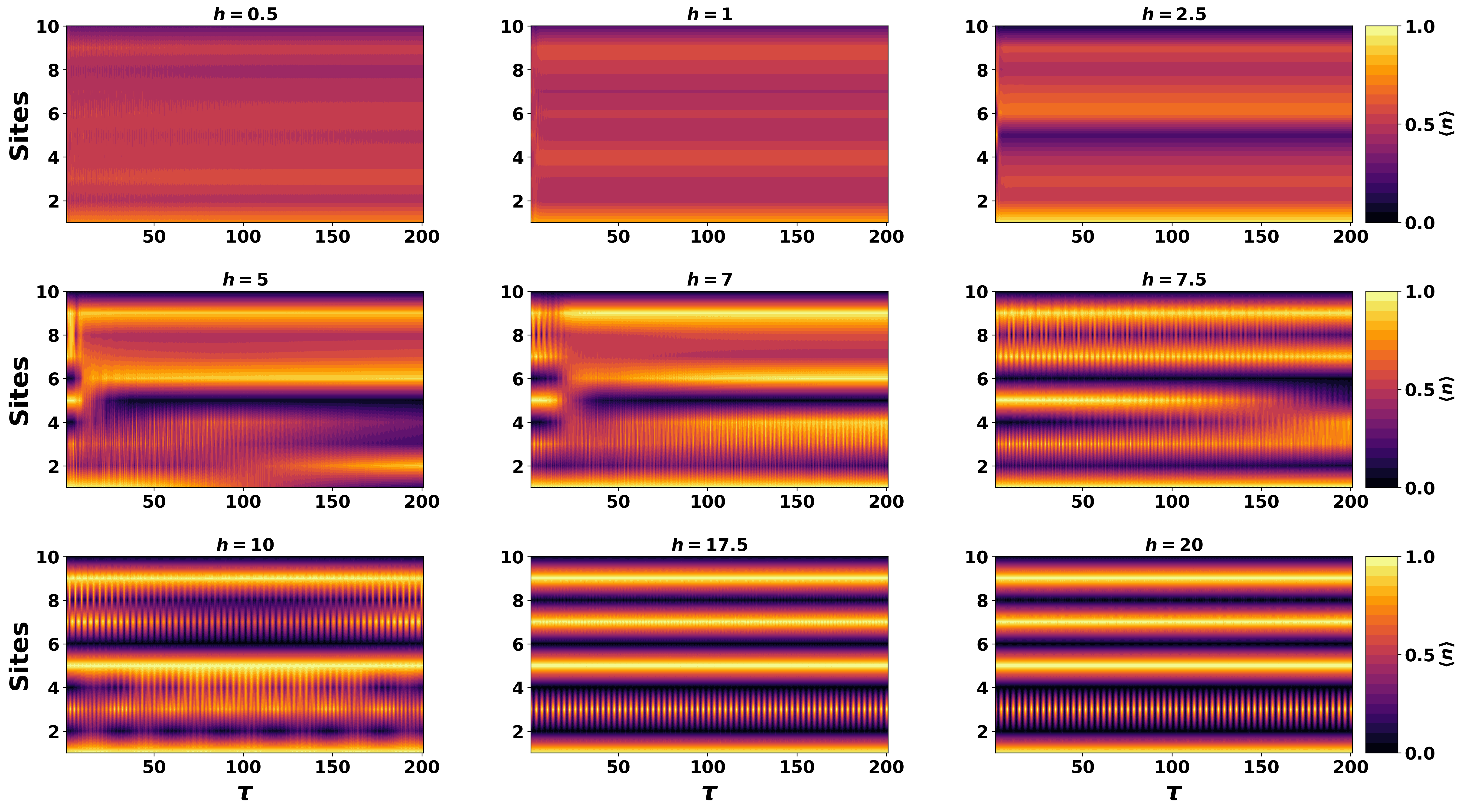}
    \caption{\textbf{Many-body localization and many-body skin effect}. We plot the average particle density $\langle n_i (\tau)\rangle$ as a function of the number of sites and the evolution time $\tau$ for different values of the disorder $h$. We observe the localization of the state at one edge for $h=0.5$, $1$, and $2.5$, which is thehallmark for the MBSE, and the emergence of MBL for larger values $h=10$, $17.5$, and $20$. For $h=5$, $7$ and $7.5$ one observes the emergence of a mixed regime characterizing the MBME. We consider $L=10$, impose OBCs and set $J=1$, $\phi=0.0$, $t/J=1/\sqrt{2}$, $\gamma/J=0.8$, and $g/J=(5+\sqrt{5})/8$.}
    \label{fig7}
\end{figure*}

Now, let us study the effect of nearest-neighbor interaction depicted in Fig.~\ref{fig6}(b, e), where we assume strong $\gamma$. For weak $h$ and under PBCs, the nearest-neighbor coupling $J$ enhances spin-spin correlations rendering the eigenstates more delocalized with weak multifractality such that, as $J$ increases, the eigenstates remain more or less extended cf. Fig.~\ref{fig6}(b). Under OBCs, for small $J$, $\langle D_2\rangle_{\mathrm{OBCs}}\approx 0.8$; the OBC--PBC gap in $\langle D_2\rangle$ is small, indicating that boundary-induced eigenstate accumulation has a limited effect on Hilbert-space localization at weak interactions. As $J$ increases, the NH asymmetry combines with interactions to create directional bias and accumulation, increasing the degree of localization, which manifests itself via a more pronounced MBSE. This leads to a drop in $\langle D_2 \rangle$ value, which serves as a fingerprint of an interaction-enhanced skin effect. Further, under strong $h$, as mentioned before, the quasi-periodic field already induces MBL, and as such $J$ has the effect of pushing $\langle D_2 \rangle$ upward and tends to delocalize eigenstates. Under PBCs \textit{``upward pushing''} is manifested by the blue peak in Fig.~\ref{fig6}(e), however, the value of $\langle D_2 \rangle $ remains relatively small ($\approx 0.1$) indicating that we still remain in the localized regime. This observation confirms the presence of localized states with weak hybridization as earlier reported in this work. The same observation is made under OBCs, where the competition between the quasi-periodic field and the weak NH parameter makes the \textit{``upward pushing''} moderate.

We now examine how $\langle D_2\rangle$ evolves with $h$ for strong
$\gamma$, probing the competition between the quasiperiodic field,
many-body interactions, and boundary-induced effects. At small $J$ [Fig.~\ref{fig6}(c)], both OBC and PBC curves follow a monotone transition from the extended to the MBL regime via the MBME with no significant gap between the two boundary conditions.
At large $J$ [Fig.~\ref{fig6}(f)], the picture is richer. At small $h$, the MBSE is dominant: wavefunction accumulation at one boundary under OBC induces a large gap between
$\langle D_2\rangle_{\mathrm{OBC}}$ and
$\langle D_2\rangle_{\mathrm{PBC}}$.
As $h$ increases, the quasiperiodic potential progressively freezes eigenstates into localized configurations, which are insensitive to boundary conditions, suppressing the MBSE and narrowing this gap.
Near the MBME, the competition between boundary-induced accumulation and field-induced localization produces a crossing, $\langle D_2\rangle_{\mathrm{OBC}} > \langle D_2\rangle_{\mathrm{PBC}}$: coherent transport responsible for the skin effect is disrupted, while PBC eigenstates remain more strongly confined by the periodic geometry, temporarily reversing the hierarchy. At large $h$, strong quasiperiodic disorder localizes all eigenstates in real space, suppressing the boundary-induced accumulation characteristic of the MBSE and causing the OBC and PBC curves to
merge, $\langle D_2\rangle_{\mathrm{OBC}} \approx
\langle D_2\rangle_{\mathrm{PBC}}$.

\section{Dynamics} \label{S5}
So far, we discussed the MBSE and the MBME from the perspective of the fractal dimension, the energy, the inverse participation ration and the angular statistics. In this subsection, we pursue a different objective: to demonstrate, as extensively discussed in Sec.~\ref{S2}, the transition from the extended to MBL regime \cite{Yutao2025ind}. In order to achieve this goal, we study the dynamics of the particle density $n_i$ (see Fig.~\ref{fig7}). Specifically, we prepare a N\'eel state $|\psi_0\rangle$ at half-filling for $L=10$, and study its quench dynamics. At time $\tau$, the state evolves to $ |\psi_{\tau}\rangle = \bm{U}|\psi_0\rangle/\sqrt{\langle \psi_0|\bm{U}^{\dagger}\bm{U}|\psi_0\rangle}$, where $\bm{U}=e^{-i\bm{H}\tau}$ is the propagator. The time evolution is carried out using Krylov subspace methods~\cite{PhysRevB.111.064203,NANDY20251,bhattacharya2023krylov,bhattacharya2022operator}. To characterize the MBSE, we compute the total average particle numbers given by $\langle n_i (\tau)\rangle=\langle \psi_{\tau}|\sum_{i=1}^L n_i|\psi_{\tau}\rangle$ at long times. 

\begin{figure*}[ht!]
     \centering
    \includegraphics[width=0.789\textwidth]{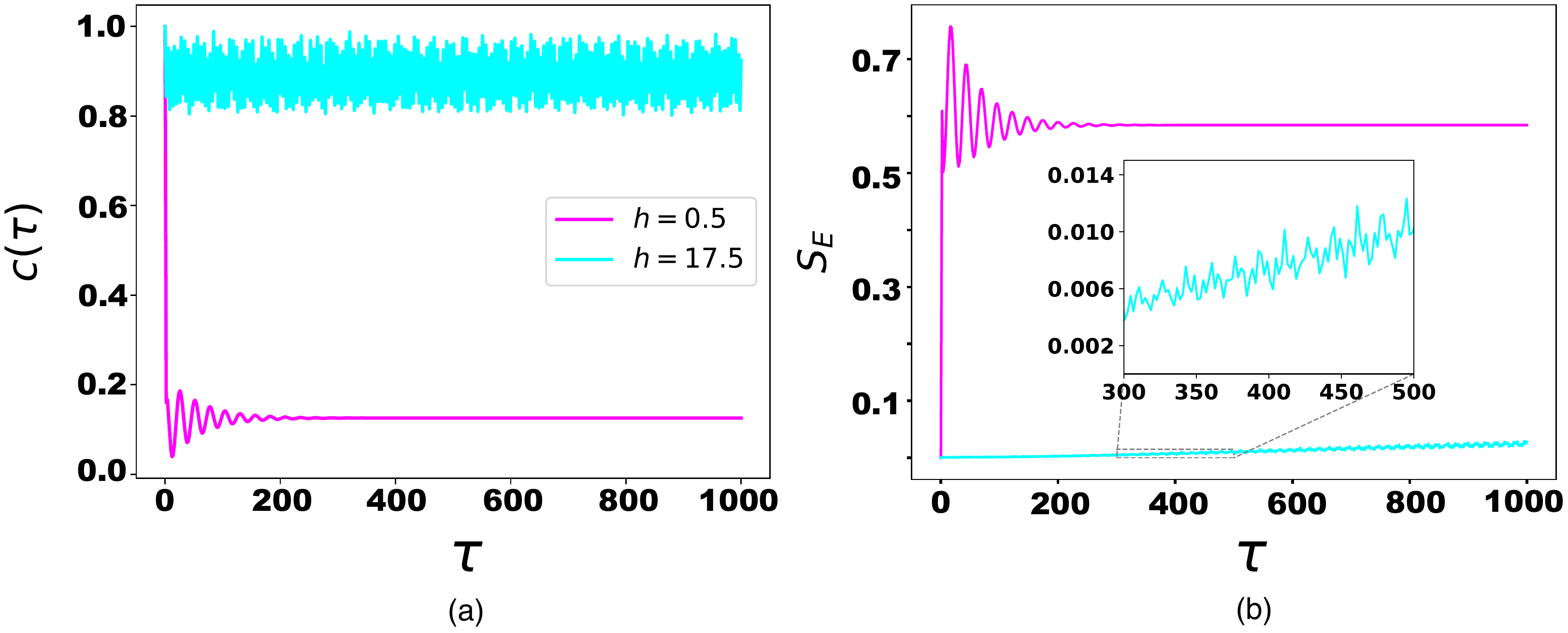}
    \caption{\textbf{Charge imbalance $c(\tau)$ and entanglement entropy $S_E$}. We plot the time-evolved \textbf{(a)} charge imbalance $c(\tau)$ and \textbf{(b)} entanglement entropy $S_E$ for $L=10$, and observe distinct features of the extended (purple) and the MBL (cyan) regimes. OBCs are considered. We set $J=1$, $t/J=1/\sqrt{2}$, $\gamma/J=0.8$, $h/J = (1+\sqrt{5})/4$, and $g/J=(5+\sqrt{5})/8$. }
    \label{fig8}%
\end{figure*}

For small disorder, ($h=0.5$ and $1$) we observe the accumulation of states at the left end of the chain (around site 1), where the average particle density $\langle n_i\rangle \approx 1$ (see the top row in Fig.~\ref{fig7}) suggesting the presence of the MBSE. In contrast when we increase the disorder to $h=10$, $17.5$, and $20$, cf. the bottom row in Fig.~\ref{fig7}, the time evolution of the averaged particle density in the system reveals a distinctive signature of MBL. Specifically, the persistent pattern of alternating bright and dark fringes reveals that the local density $\langle n_i (\tau)\rangle$ remains close to $0$ on odd sites and approaches $1$ on even sites. This indicates that the system retains memory of its initial state over long times. Indeed, in the MBL phase, local excitations fail to propagate freely because of strong  quasi-periodic fields and interactions, causing the system to become dynamically localized in its many-body Hilbert space. Consequently, the initial density-wave pattern remains frozen or only weakly perturbed, in stark contrast to the extended regime, where such a structure quickly dissolves due to the delocalization and scrambling of information as discussed before. The persistent N\'eel-like density profile observed is a direct evidence of the emergence of MBL.

It is instructive to note that in the intermediate regime for $h=5$, $7$, and $7.5$, as shown in the middle row in Fig.~\ref{fig7}, one observes the transition from the extended to the MBL regime. For example, we notice the delocalization of states in the first five sites, while the state is localized on the last five sites. The transition region within which these two regimes coexist corresponds to the region in which we identified a MBME to exist.

\subsection{Charge imbalance and entanglement entropy}

We now delve deeper into this discussion by examining quantities such as charge imbalance and entanglement entropy, focusing on the role of quasi-periodic disorder and the behavior in the extended and MBL regimes. These quantities are well-established diagnostics in Hermitian settings, where they are experimentally measurable~\cite{exp1,exp2}, and we use them here to further support our theoretical results.

First, we measure the disorder-induced localization through the charge imbalance $c(\tau)$ given by
\begin{align}\label{E5}
    c(\tau)= \frac{1}{L} \sum_i (-1)^i \langle n_i (\tau)\rangle.
\end{align}
Figure~\ref{fig8}(a) shows the disorder-averaged charge imbalance $c(\tau)$ for representative values of $h$. In the extended phase ($h=0.5$, purple), $c(\tau)$ decays to zero, signalling complete relaxation of the initial N\'eel order and loss of memory of the initial state. In the MBL phase ($h=17.5$, cyan), $c(\tau)$ saturates to a finite nonzero value, reflecting long-time memory retention irrespective of $J$ and $\gamma$~\cite{Suthar2025Bound}.

Next, we investigate the growth of the entanglement entropy $S_E(\tau)$ define as 
\begin{align}
    S_E(\tau)=-Tr[\rho_A(\tau)\ln\rho_A(\tau)],
\end{align}\label{E6}
\hspace{-0.3cm} where $\rho_A(\tau)$ is the reduced density matrix of one half of the system defined in terms of right eigenstates. Fig.~\ref{fig8}(b) shows that in the extended regime (purple curve, $h=0.5$), there is a rapid growth and saturation of the entanglement entropy at large values. This is the expected behavior, and is analogous to the behavior in Hermitian systems. For $h=17.5$, we observe a slow  growth of the entanglement entropy as time passes (cyan curve), which suggests that we are in the MBL regime.

\section{Conclusion}\label{S6}

In summary, the collective behavior of non-Hermiticity, many-body interactions, 
and (quasi-periodic) disorder in a single unified setting represents one of the most fertile yet least explored frontiers of open quantum many-body physics. By studying an interacting NH spin chain subject to a quasi-periodic longitudinal field, we have shown that these three ingredients conspire to produce a remarkably structured phase diagram, governed 
by the competition between boundary-induced eigenstate accumulation (MBSE), interaction-enhanced localization (MBL), and the multifractal scaling that emerges at their interface. Central to our findings is 
the discovery of a \textit{D-shaped} MBME 
that simultaneously delineates the extended and MBL phases, and marks the crossover between regimes dominated by the MBSE (under OBCs) and those frozen by quasi-periodic disorder. Thisestablishes the \textit{D-shaped} MBME as a robust feature of the extended-to-localized transition in open quantum many-body systems.

To substantiate these claims, we employed a comprehensive set of complementary diagnostics. The scaling of the $q$-dependent fractal dimensions $D_q$ with the participation entropy confirmed the multifractal behavior, while in the weak-disorder (delocalized) regime, a systematic drift of the particle density toward the left boundary under OBC---yielding a finite-density accumulation at half filling---
provided a direct real-space signature of the MBSE. Upon increasing the quasi-periodic field strength, the system transitions from the 
delocalized phase into an MBL phase, yet this transition is not sharp: rather, it is mediated by a broad intermediate region in which localized and delocalized eigenstates coexist within the same spectrum and $\langle D_2 \rangle$ is widely spread---the defining spectral fingerprint of the \textit{``D-shaped''} MBME. This unified picture 
was consistently corroborated by three independent probes: the MIPR, the complex level-spacing ratio, and dynamical observables, namely, charge imbalance, entanglement entropy, and wave-packet dynamics, all of which all identify the same phase boundaries.

Our results open several promising directions for future investigation. On the theoretical side, constructing NH analogues of local integrals of motion (LIOMs) would clarify whether the emergent 
integrability of the MBL phase survives the introduction of non-Hermiticity. In particular, for strong quasiperiodic disorder, the system retains memory of its initial state, a hallmark of MBL widely understood through the framework of LIOMs~\cite{Singh2021local, Bertoni2023local, lu2024measuring, shtanko2025uncovering}. In this picture, an extensive set of emergent quasi-local conserved quantities $\{\eta_i^z\}$, each localized near site $i$, underlies the failure 
of thermalization despite interactions with the Hamiltonian admitting an approximate representation satisfying $[H,\eta_i^z]\approx 0$. While most naturally formulated for random disorder, an analogous construction is expected to hold for quasiperiodic potentials where localization arises from incommensurability. An analogous LIOM structure in the NH setting, defined with respect to the biorthogonal eigenbasis, would naturally explain the logarithmic entanglement growth in the MBL phase: transport is frozen, yet phase information spreads slowly via dephasing, generating a logarithmic entanglement buildup without violating localization. Crucially, since all eigenstates are exponentially localized in the MBL phase, the MBSE is naturally suppressed, rendering the LIOM picture robust even in the NH setting. A systematic derivation of LIOMs in NH systems remains an important open direction.

Besides these considerations, it would be illuminating to examine whether the \textit{D-shaped} MBME persists under random Gaussian disorder, which would establish it as a feature of the ergodic-to-localized transition rather than an artifact of the quasi-periodic potential. More broadly, extending this framework to higher-dimensional NH systems or to Markovian open quantum systems described by a Lindblad master equation remains an important open challenge.

Finally, the computation of the key observables identified here, i.e., density imbalance, entanglement growth, and 
wave-packet evolution,suggests that the MBSE-to-MBL crossover and its \textit{D-shaped} boundary could in principle be directly probed in ultracold atomic gases, photonic lattices, and digital quantum processors operating in the NH regime.

\section{Acknowledgments}
We acknowledge funding from the Max Planck Society Lise Meitner Excellence Program~\mbox{2.0}. F.K.K. also acknowledges funding from the European Union's ERC Starting Grant ``NTopQuant'' (101116680). Views and opinions expressed are, however, those of the authors only and do not necessarily reflect those of the European Union or the European Research Council (ERC). Neither the European Union nor the granting authority can be held responsible for them.

\section{Data Availability}

The data and codes that support the findings of this study are available from the corresponding author upon reasonable request.



\bibliography{bibliography}

\end{document}